\documentclass[english,british]{paper}
\usepackage[T1]{fontenc}
\usepackage[latin9]{inputenc}
\usepackage{color}
\usepackage{babel}
\usepackage{amsmath}
\usepackage{amssymb}
\usepackage{graphicx}
\usepackage{esint}
\usepackage[unicode=true,pdfusetitle,
 bookmarks=true,bookmarksnumbered=false,bookmarksopen=false,
 breaklinks=false,pdfborder={0 0 0},backref=false,colorlinks=true]
 {hyperref}
\hypersetup{
 linkcolor=blue,citecolor=blue}

\makeatletter
\usepackage{babel}
\usepackage{babel}
\usepackage{babel}
\renewcommand{\[}{\begin{equation}}

\renewcommand{\]}{\end{equation}}

\makeatother

\begin{document}

\title{Action-at-a-distance electrodynamics in Quasi-steady-state cosmology}

\author{Kaustubh Sudhir Deshpande$^{1,2}$ \\
 (Communicated by Prof. J. V. Narlikar) }

\maketitle
\begin{small}

\noindent $^{1}$Indian Institute of Science Education and Research
(IISER), Pashan, Pune 411 021, India. \\
 $^{2}$Current address: Harish-Chandra Research Institute, Chhatnag
Road, Jhusi, Allahabad 211 019, India. \\
 Email address: k.deshpande@alumni.iiserpune.ac.in \end{small}
\begin{keywords}
Action-at-a-distance electrodynamics; Quasi-steady-state cosmology;
Retarded and advanced interactions; Wheeler-Feynman absorber theory\end{keywords}
\begin{abstract}
Action-at-a-distance electrodynamics - alternative approach to field
theory - can be extended to cosmological models using conformal symmetry.
An advantage of this is that the origin of arrow of time in electromagnetism
can be attributed to the cosmological structure. Different cosmological
models can be investigated, based on Wheeler-Feynman absorber theory,
and only those models can be considered viable for our universe which
have net full retarded electromagnetic interactions i.e. forward direction
of time. This work evaluates quasi-steady-state model and demonstrates
that it admits full retarded and not advanced solution. Thus QSSC
satisfies this necessary condition for a correct cosmological model,
based on action-at-a-distance formulation.
\end{abstract}

\section{Introduction}

Newton's law of gravitation and Coulomb's law for electrical charges,
one of the very first laws of Theoretical Physics, assumed instantaneous
action-at-a-distance between particles. The gravitational and electrical
effects due to masses and charges respectively, were assumed to travel
at infinite speed in these laws. The experiments in electrodynamics,
however, demonstrated that Coulomb's law was inadequate to explain
their results. Gauss suggested action-at-a-distance propagating at
a finite speed (speed of light) in his letter to Weber in 1845 \cite{Gauss ADE}.
However, this did not get immediately formulated. Instead, Maxwell
developed classical field theory of electrodynamics which has effects
propagating at the speed of light. This was also found to be consistent
with special relativity (which discarded instantaneous action-at-a-distance
of Newton's and Coulomb's laws). Maxwell field theory can be described by the following
relativistically invariant action,
\[
S=-\sum_{a}\int m_{a}ds_{a}-\frac{1}{16\pi}\int F_{ik}F^{ik}d^{4}x-\sum_{a}\int e_{a}A_{i}dx_{a}^{i}
\]

\noindent where $F_{ik}$ is the field, with infinite degrees of freedom,
defined in terms of the 4-potential $\left(A_{i}\right)$ as $F_{ik}=\left(A_{k;i}-A_{i;k}\right)$.
The particles (labelled by $a,b,...$) do not interact directly with
each other but interact through their coupling with the field (described
by the third term in $S$).

In early 20th century, Schwarzschild \cite{Schwarzschild ADE}, Tetrode
\cite{Tetrode ADE} and Fokker \cite{Fokker ADE 1,Fokker ADE 2,Fokker ADE 3}
developed a relativistically invariant action-at-a-distance theory.
This partially found the answer to Gauss's problem. This theory can
be described by the Fokker action which is given as follows.
\[
S=-\sum_{a}\int m_{a}ds_{a}-\sum_{a<b}\int e_{a}e_{b}\:\delta(s_{AB}^{2})\:\eta_{ik}dx_{a}^{i}dx_{b}^{k}
\]

The first term is the same inertial term as in the field theoretic
action. The second term represents electromagnetic interactions between
two different particles $a,b$ connected by a light cone%
\footnote{implied by $s_{AB}^{2}=\eta_{ij}(x_{a}^{i}-x_{b}^{i})(x_{a}^{j}-x_{b}^{j})$
being zero. $A$, $B$ are typical points on worldlines of the two
particles $a$, $b$.%
}, thus preserving relativistic invariance. This action with the following
definitions of direct particle potentials $\left(A_{i}^{(b)}\right)$
and fields $\left(F_{ik}^{(b)}\right)$,
\[
A_{i}^{(b)}(X)=e_{b}\int\delta(s_{XB}^{2})\eta_{ik}dx_{b}^{k},\quad F_{ik}^{(b)}=A_{k;i}^{(b)}-A_{i;k}^{(b)}
\]

\noindent gives exactly Maxwell-like equations for $F_{ik}^{(b)}$
and Lorentz-like equations of motion for the particles \cite{Hoyle Narlikar review 1995}.
This formulation hence resembles and seems to provide an alternative
to the Maxwell's field theory.

Action-at-a-distance in electrodynamics was provided a paradigm by
Wheeler and Feynman in 1945 \cite{Wheeler Feynman 1945} through their
absorber theory of radiation. This theory, formulated in static and
flat universe, uses advanced absorber response from the entire universe
as the origin for radiation reaction but, being time-symmetric in
nature, allows for both retarded and advanced net interactions as
consistent solutions. Wheeler and Feynman broke this symmetry in favour
of the former by making an appeal to Statistical mechanics.

Extension of this formulation to expanding cosmological models (using
conformal invariance of electromagnetism and conformal flatness of
cosmological models) was done by Hogarth \cite{Hogarth 1962} and
more generally by Hoyle-Narlikar \cite{Hoyle Narlikar about WF theory in cosmology 1964}.
Self-consistency of net advanced and retarded interactions in these
models can be investigated by evaluating Wheeler-Feynman absorption
integrals. Only those models can be considered to be viable which
have only net retarded interactions (as observed in nature). It was
found by Hoyle-Narlikar that steady-state model satisfies this criterion
while Friedman models do not \cite{Hoyle Narlikar about WF theory in cosmology 1964}.

Action-at-a-distance electrodynamics has to be first formulated for
a generalized Riemannian space-time, before appyling it to cosmological
space-times using conformal symmetry. This has been formulated by
Hoyle-Narlikar \cite{Hoyle Narlikar about WF theory in cosmology 1964,Hoyle Narlikar-Action at a distance-book}
by generalizing Fokker action to curved space-times using Green's
functions for wave equation.

Here we evaluate Wheeler-Feynman (WF) absorption integrals in Quasi-steady-state
cosmology (QSSC) to investigate the self-consistency for net retarded
and advanced electromagnetic interactions.

\section{Quasi-steady-state cosmology}

QSSC, based on the steady-state cosmology, is an alternative model
to the standard cosmology. It was proposed by Hoyle, Burbidge and
Narlikar in 1993 \cite{QSSC_Hoyle Burbidge and Narlikar}. This is
cosmological solution for Machian theory of gravity proposed by Hoyle
and Narlikar in 1964 \cite{Machian gravity_Hoyle-JVN_1964} with periodic
creation of matter. The line element of QSSC can be represented as,

\begin{eqnarray*}
ds^{2} & = & d\tau^{2}-S^{2}(\tau)\left[dr^{2}+r^{2}\left(d\theta^{2}+\sin^{2}\theta d\phi^{2}\right)\right]\\
 & = & \Omega^{2}(t)\left[dt^{2}-dr^{2}-r^{2}\left(d\theta^{2}+\sin^{2}\theta d\phi^{2}\right)\right]
\end{eqnarray*}

The scale factor%
\footnote{$S(\tau)=\textrm{scale factor},\;\Omega(t)=\textrm{conformal factor}$%
} of QSSC has the following form which is expanding exponentially over
a large time scale $(P)$ and oscillating over a much shorter time
scale $(Q)$. Typically, $Q=50$ gigayears, $P=20Q$ \cite{Narlikar_Introduction to cosmology}.
\[
S(\tau)=\exp\left(\frac{\tau}{P}\right)\left[1+\eta\cos\theta\left(\tau\right)\right]\quad:-\infty\leq\tau\leq\infty
\]

\begin{figure}[h]
\includegraphics[scale=0.5]{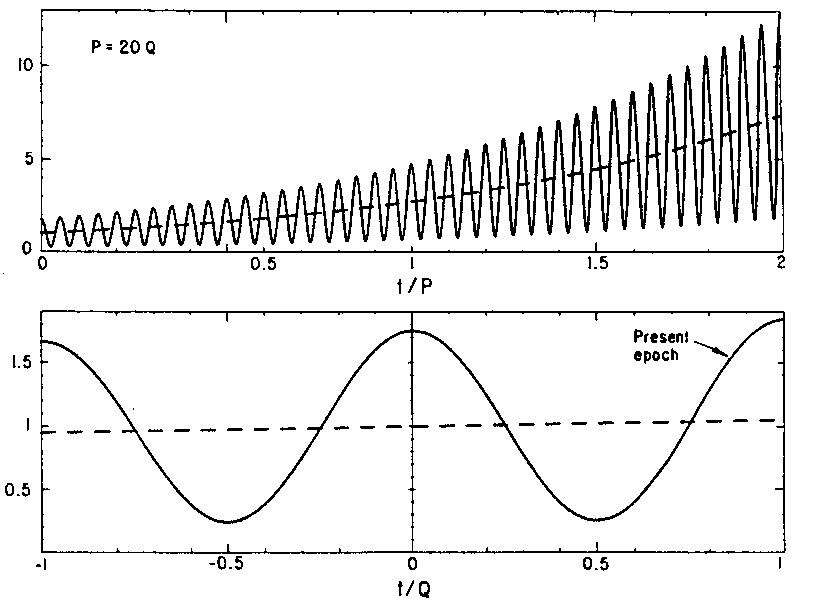}

\caption{Scale factor for QSSC\label{fig:QSSC-Scale-factor}: $S(\tau)$ expands
exponentially over a time scale $P$ and oscillates over a much smaller
time scale $Q$. The present epoch ($\tau_{0}$) denotes decelerating expansion
which is supported by reinterpretation of magnitude-redshift relation
for supernovae in QSSC. Intergalactic dust in QSSC makes supernovae
dimmer and thermalizes starlight to produce CMBR. \cite{QSSC_arxiv review_Kragh,QSSC deceleration_intergalactic dust,Hoyle Burbidge Narlikar_Different approach to cosmology}}
\end{figure}

The function $\theta\left(\tau\right)$ can be simplified and in an
approximation,
\[
S(\tau)=\exp\left(\frac{\tau}{P}\right)\left[1+\eta\cos\left(\frac{2\pi\tau}{Q}\right)\right]
\]

\noindent where $P>>Q$, $\eta=$ constant with $0<\eta<1$. See Fig.\ref{fig:QSSC-Scale-factor}.

The number density $(N)$ of particles oscillates between $N_{min}$
and $N_{max}$ during a cycle of time period $Q$ but the average
density remains constant due to periodic creation of matter at particular
time epochs \cite{Narlikar_Introduction to cosmology}.

With the following series of inequalities, we get the bounds on conformal
factor, $\Omega(t)$.

\begin{center}
\begin{equation}
\exp\left(\frac{\tau}{P}\right)\left(1-\eta\right)\leq S(\tau)\leq\exp\left(\frac{\tau}{P}\right)\left(1+\eta\right)\label{eq:Inequality on scale factor}
\end{equation}

\par\end{center}

\begin{center}
\[
\int_{\infty}^{\tau}d\tau\frac{e^{-\tau/P}}{\left(1-\eta\right)}\leq\left[t=\int_{\infty}^{\tau}\frac{d\tau}{S(\tau)}\right]\leq\int_{\infty}^{\tau}d\tau\frac{e^{-\tau/P}}{\left(1+\eta\right)}\quad:-\infty\leq t\leq0
\]

\par\end{center}

\[
-\frac{P}{t}\left(\frac{1-\eta}{1+\eta}\right)\leq e^{\tau/P}\left(1-\eta\right)\leq\Omega(t)\leq e^{\tau/P}\left(1+\eta\right)\leq-\frac{P}{t}\left(\frac{1+\eta}{1-\eta}\right)
\]

\begin{center}
\begin{equation}
-\frac{P}{t}\left(\frac{1-\eta}{1+\eta}\right)\leq\Omega(t)\leq-\frac{P}{t}\left(\frac{1+\eta}{1-\eta}\right)\label{eq:Inequality on conformal factor}
\end{equation}

\par\end{center}

Thus (\ref{eq:Inequality on conformal factor}) denotes the bounds
on conformal factor $\Omega(t)$. It is important to note that the
oscillatory microstructure in the scale factor $S(\tau)$ is not being
ignored or approximated here. We are instead considering bounds on
the functional forms, in an appropriate way. Hence the results obtained
in the next section (regarding WF absorption integrals) are exact
and not approximate.

\section{Evaluation of WF absorption integrals: retarded and advanced waves}

In this section we evaluate WF absorption integrals ($I_{R}$ and
$I_{A}$) in QSSC, to check self-consistency of retarded ($R$) and
advanced ($A$) solutions. We are using Hoyle-Narlikar approach \cite{Hoyle Narlikar about WF theory in cosmology 1964} for this evaluation, with radiation reaction provided by absorber response from the entire universe. The conditions required for this are given
by the divergences of the following integrals%
\footnote{for spatially flat universe%
}.

\begin{equation}
I_{R}=\int_{\textrm{future}}k(r)dr\rightarrow-\infty,\; I_{A}=\int_{\textrm{past}}k(r)dr\rightarrow+\infty\quad;k=\textrm{absorption coefficient}\label{eq:WF absorption integrals}
\end{equation}

Cosmological shift in the frequency of wave%
\footnote{The range and constraints on coordinates - for retarded wave: $r=t-t_{0}$,
$t:t_{0}\rightarrow0,\, r:0\rightarrow-t_{0}$; for advanced wave:
$r=t_{0}-t$, $t:t_{0}\rightarrow-\infty,\, r:0\rightarrow\infty$. %
} is given by,
\begin{equation}
\omega\propto S^{-1}\propto\Omega^{-1}\label{eq:Frequecy vs. Scale factor}
\end{equation}

\[
\omega_{R}=\omega_{0}\left(\Omega^{-1}\left(t_{0}+r\right)\right),\quad\omega_{A}=\omega_{0}\left(\Omega^{-1}\left(t_{0}-r\right)\right)
\]

Using (\ref{eq:Inequality on conformal factor}), $\omega_{R}$ and
$\omega_{A}$ satisfy the following inequalities.

\begin{equation}
-\omega_{0}\frac{\left(t_{0}+r\right)}{P}\left(\frac{1-\eta}{1+\eta}\right)\leq\omega_{R}\leq-\omega_{0}\frac{\left(t_{0}+r\right)}{P}\left(\frac{1+\eta}{1-\eta}\right)\label{eq:Inequality on retarded wave frequency}
\end{equation}

\begin{equation}
-\omega_{0}\frac{\left(t_{0}-r\right)}{P}\left(\frac{1-\eta}{1+\eta}\right)\leq\omega_{A}\leq-\omega_{0}\frac{\left(t_{0}-r\right)}{P}\left(\frac{1+\eta}{1-\eta}\right)\label{eq:Inequality on advanced wave frequency}
\end{equation}

The refractive index satisfies the following expressions in the respective
limits%
\footnote{\selectlanguage{english}%
$\omega\rightarrow0$ and $\omega\rightarrow\infty$ limits correspond
to future and past infinities respectively.\selectlanguage{british}%
}\cite{Hoyle Narlikar about WF theory in cosmology 1964}.

\[
1-(n-ik)^{2}=\begin{cases}
\frac{4\pi Ne^{2}}{m\omega^{2}}\left[1-\frac{2ie^{2}}{3m}\omega+O(\omega^{2})\right] & ;\omega\rightarrow0\\
\frac{4\pi Ne^{2}}{m\omega^{2}}\left[1+O\left(\frac{1}{\omega}\right)\right] & ;\omega\rightarrow\infty
\end{cases}
\]

Hence absorption coefficient $(k)$ satisfies,

\begin{equation}
k_{R}\sim-\frac{\sqrt{N}}{\omega},\quad k_{A}\sim\frac{N}{\omega^{3}}\label{eq:Dependence of k on frequency}
\end{equation}

Using (\ref{eq:Inequality on retarded wave frequency}) for retarded
solution and the fact that $N_{min}\leq N\leq N_{max}$ for QSSC,

\[
\frac{\sqrt{N_{max}}}{\omega_{0}}P\left(\frac{1+\eta}{1-\eta}\right)\frac{1}{\left(t_{0}+r\right)}\leq\left[k_{R}\sim-\frac{\sqrt{N}}{\omega}\right]\leq\frac{\sqrt{N_{min}}}{\omega_{0}}P\left(\frac{1-\eta}{1+\eta}\right)\frac{1}{\left(t_{0}+r\right)}
\]

\[
\frac{C_{2}}{t_{0}+r}\leq k_{R}\leq\frac{C_{1}}{t_{0}+r}
\]

The future absorption integral $\left(I_{R}\right)$ thus satisfies%
\footnote{$C_{1}$ and $C_{2}$ are positive, finite constants. $C_{1}=\frac{\sqrt{N_{min}}}{\omega_{0}}P\left(\frac{1-\eta}{1+\eta}\right),C_{2}=\frac{\sqrt{N_{max}}}{\omega_{0}}P\left(\frac{1+\eta}{1-\eta}\right)$%
},

\[
C_{2}\int_{0}^{-t_{0}}\frac{dr}{t_{0}+r}\leq\left[I_{R}=\int_{\textrm{future}}k_{R}(r)dr\right]\leq C_{1}\int_{0}^{-t_{0}}\frac{dr}{t_{0}+r}
\]

As integrals on either side of $I_{R}$ in the above inequality diverge
to $-\infty$, $I_{R}$ also diverges to $-\infty$. Thus QSSC satisfies
the condition for $I_{R}$ (see (\ref{eq:WF absorption integrals}))
and admits self-consistent retarded solution.

Similarly using (\ref{eq:Inequality on advanced wave frequency})
for advanced solution,

\[
-\frac{N_{min}}{\omega_{0}^{3}}P^{3}\left(\frac{1-\eta}{1+\eta}\right)^{3}\frac{1}{\left(t_{0}-r\right)^{3}}\leq\left[k_{A}\sim\frac{N}{\omega^{3}}\right]\leq-\frac{N_{max}}{\omega_{0}^{3}}P^{3}\left(\frac{1+\eta}{1-\eta}\right)^{3}\frac{1}{\left(t_{0}-r\right)^{3}}
\]

\[
\frac{D_{2}}{(r-t_{0})^{3}}\leq k_{A}\leq\frac{D_{1}}{(r-t_{0})^{3}}
\]

The past absorption integral $\left(I_{A}\right)$ satisfies%
\footnote{$D_{1}$ and $D_{2}$ are positive, finite constants. $D_{1}=\frac{N_{max}}{\omega_{0}^{3}}P^{3}\left(\frac{1+\eta}{1-\eta}\right)^{3}$,
$D_{2}=\frac{N_{min}}{\omega_{0}^{3}}P^{3}\left(\frac{1-\eta}{1+\eta}\right)^{3}$%
},

\[
D_{2}\int_{0}^{\infty}\frac{dr}{(r-t_{0})^{3}}\leq\left[I_{A}=\int_{\textrm{past}}k_{A}(r)dr\right]\leq D_{1}\int_{0}^{\infty}\frac{dr}{(r-t_{0})^{3}}
\]

\[
\Rightarrow\frac{D_{2}}{2t_{0}^{2}}\leq I_{A}\leq\frac{D_{1}}{2t_{0}^{2}}
\]

This shows that $I_{A}$ takes a finite value and hence QSSC does
not admit self-consistent advanced solution.

It is important to note that the absorption property derived here
(for both future and past absorption) is a robust result. It is independent
of where the observer is located. (This is also evident from the fact
that nowhere in the calculation the value $\tau_{0}$, or local
qualitative behaviour related to its position, is required). Hence
the same result applies to observers located at epochs corresponding
to deep crest or trough of the oscillation. This is because the underlying
theory being global (and not local) in nature, one has to consider
absorber response from the entire universe. This amounts to evaluating
WF absorption integrals $I_{R}$ and $I_{A}$ till future and past
infinities, respectively. Thus local variations or position of the
observer do not affect the results in any significant way.

\section{Additional supplementary calculation of WF integrals}

In order to support the note made in the last paragraph of previous
section, we present an additional analysis in this section to demonstrate
that the self-consistency conditions in QSSC are independent of $\frac{P}{Q}$
or local position of the observer (i.e. in crest or trough of the
oscillation).

Null, radial light ray in QSSC satisfies the following.

\[
\frac{dr}{d\tau}=\frac{1}{S(\tau)},\qquad S(\tau)=\exp\left(\frac{\tau}{P}\right)\left[1+\eta\cos\left(\frac{2\pi\tau}{Q}\right)\right]
\]

Using (\ref{eq:Frequecy vs. Scale factor}), we have $\omega(\tau)=\frac{\omega_{0}}{S(\tau)}$.

Now, using the dependence of $k_{R}$ and $k_{A}$ on $\omega$ as
given in (\ref{eq:Dependence of k on frequency}), we obtain the following
results for $I_{R}$ and $I_{A}$.

\[
I_{R}=\int_{\textrm{future}}k_{R}dr=\int_{\tau_{0}}^{\infty}k_{R}\frac{dr}{d\tau}d\tau\sim-\int_{\tau_{0}}^{\infty}\sqrt{N}d\tau
\]

\[
-\sqrt{N_{max}}\int_{\tau_{0}}^{\infty}d\tau\leq\left[I_{R}\sim-\int_{\tau_{0}}^{\infty}\sqrt{N}d\tau\right]\leq-\sqrt{N_{min}}\int_{\tau_{0}}^{\infty}d\tau
\]

Hence it can be seen that, $I_{R}\rightarrow-\infty$, completely
independent of the value $\frac{P}{Q}$ or $\tau_{0}$%
\footnote{$\tau_{0}$ is the present epoch i.e. epoch of the position of the
observer.%
}.

\[
I_{A}=\int_{\textrm{past}}k_{A}dr=\int_{\tau_{0}}^{-\infty}k_{A}\frac{dr}{d\tau}d\tau\sim\int_{\tau_{0}}^{-\infty}NS^{2}(\tau)d\tau
\]

\[
N_{min}\int_{\tau_{0}}^{-\infty}S^{2}(\tau)d\tau\leq\left[I_{A}\sim\int_{\tau_{0}}^{-\infty}NS^{2}(\tau)d\tau\right]\leq N_{max}\int_{\tau_{0}}^{-\infty}S^{2}(\tau)d\tau
\]

\begin{align*}
\Lambda\equiv\int_{\tau_{0}}^{-\infty}S^{2}(\tau)d\tau & =\frac{P}{4}e^{2\tau_{0}/P}[\frac{1}{\left(4\pi^{2}P^{2}+Q^{2}\right)}\left\{ 2\pi\eta^{2}PQ\sin\left(\frac{4\pi\tau_{0}}{Q}\right)+\eta^{2}Q^{2}\cos\left(\frac{4\pi\tau_{0}}{Q}\right)\right\} \\
 & +\frac{1}{\left(\pi^{2}P^{2}+Q^{2}\right)}\left\{ 4\pi\eta PQ\sin\left(\frac{2\pi\tau_{0}}{Q}\right)+4\eta Q^{2}\cos\left(\frac{2\pi\tau_{0}}{Q}\right)\right\} +\eta^{2}+2]
\end{align*}

Considering $P=n\frac{Q}{2\pi}$,

\begin{align*}
\Lambda & =\frac{Q}{2\pi}ne^{4\pi\tau_{0}/nQ}[\frac{\eta^{2}}{1+n^{2}}\left\{ n\sin\left(\frac{4\pi\tau_{0}}{Q}\right)+\cos\left(\frac{4\pi\tau_{0}}{Q}\right)\right\} \\
 & +\frac{8\eta}{4+n^{2}}\left\{ n\sin\left(\frac{2\pi\tau_{0}}{Q}\right)+2\cos\left(\frac{2\pi\tau_{0}}{Q}\right)\right\} +\eta^{2}+2]
\end{align*}

In the case of pure oscillations ($n\rightarrow\infty$ limit), i.e. with $\exp\left(\frac{\tau}{P}\right)$ term
absent in $S(\tau)$, $\Lambda\rightarrow\infty$ and hence $I_{A}\rightarrow\infty$.
Thus since both $I_{R}$ and $I_{A}$ diverge in this case, both future
and past absorbers are perfect. This model of purely oscillatory universe
thus has ambiguous outcome for causality and an external criterion
(e.g. statistical mechanics, as used in \cite{Wheeler Feynman 1945}
for flat space-time) would be required to decide the direction of
time.

In the pure expansion case ($Q\rightarrow\infty$ limit), i.e. with
the oscillatory term absent in $S(\tau)$, $\Lambda=$\foreignlanguage{english}{
finite and hence $I_{A}=$ finite. This limit corresponds to Steady-state
cosmological model and admits self-consistent retarded but not advanced
solution. This result agrees with that demonstrated in \cite{Hoyle Narlikar about WF theory in cosmology 1964,Hoyle Narlikar-Action at a distance-book}.}

In QSSC model, however, the time scales $P$ and $Q$ assume positive,
finite and non-zero values. Hence the value of $n$ is large%
\footnote{as $P>>Q$%
} but finite. For any finite non-zero value of $n$, $\Lambda$ and
hence $I_{A}$ converge to finite values. Hence QSSC satisfies self-consistency
condition for retarded but not advanced solution.

It can also be noted that all the above results are unchanged for
any value of $\frac{\tau_{0}}{Q}$, i.e. position of the observer in oscillation cycle. Hence the self-consistency conditions
are the same for observers located in crests or troughs of the oscillation.

\section{Conclusion}

The above calculations show that in QSSC, $I_{R}=\int_{\textrm{future}}k(r)dr\rightarrow-\infty$
and $I_{A}=\int_{\textrm{past}}k(r)dr\rightarrow\textrm{finite value}$.
This thus concludes, in a self consistent way, that QSSC assumes only
net full retarded electromagnetic interactions. This causality is
applicable at all time epochs, as illustrated in the previous section.
QSSC is thus a viable cosmological model according to action-at-a-distance
formulation. Though this cannot be the sufficient criterion to decide
for the correct cosmological model, it certainly is a necessary criterion,
according to this approach.

Also, this has interesting implications on the origin of arrow of
time. The choice of the direction of time is \textit{ad hoc} in field
theory i.e. the retarded solution is chosen arbitrarily over the advanced
one. However in action-at-a-distance formulation, origin of time asymmetry
can be attributed to the large scale structure of the universe. The
universe has such a cosmological stucture that it provides the correct
absorber response to produce net retarded interactions, thus fixing
the arrow of time. Quasi-steady-state model is a suitable candidate
for such a cosmological structure, as shown in the present work.

\paragraph{Acknowledgements}

I would like to express deep gratitude towards Prof. J. V Narlikar
for suggesting this problem to work on and his constant guidance through
discussions. I am grateful towards IUCAA Pune and IISER Pune
for providing access to the facilities at these institutes. I would also like to thank the (anonymous) journal referee for providing useful comments and suggestions.


\begin{thebibliography}{10}
\bibitem{Gauss ADE}C F Gauss, Werke 5, 629 (1867)

\bibitem{Schwarzschild ADE}K Schwarzschild, Gottinger Nachrichten,
128, 132 (1903)

\bibitem{Tetrode ADE}H Tetrode, Z. Phys. 10, 317 (1922)

\bibitem{Fokker ADE 1}A D Fokker, Z. Phys. 58, 386 (1929)

\bibitem{Fokker ADE 2}A D Fokker, Physica 9, 33 (1929)

\bibitem{Fokker ADE 3}A D Fokker, Physica 12, 145 (1932)

\bibitem{Hoyle Narlikar review 1995}F Hoyle and J V Narlikar, Rev.
Mod. Phys., Vol. 67, No. 1, 113 (1995)

\bibitem{Wheeler Feynman 1945}J A Wheeler and R P Feynman, Rev. Mod.
Phy., 17, 157 (1945)

\bibitem{Hogarth 1962}J E Hogarth, Proc. Roy. Soc. A, 267, 365 (1962)

\bibitem{Hoyle Narlikar about WF theory in cosmology 1964}F Hoyle
and J V Narlikar, Proc. Roy. Soc. A, 277, 1 (1964)

\bibitem{Hoyle Narlikar-Action at a distance-book}F Hoyle and J V
Narlikar, Action at a distance in Physics and Cosmology (W H Freeman
and Company, San Francisco, 1974), Chap. 5, p. 85

\bibitem{QSSC_Hoyle Burbidge and Narlikar}F Hoyle, G Burbidge, J
V Narlikar, The Astrophy. J., 410, 437 (1993)

\bibitem{Machian gravity_Hoyle-JVN_1964}F Hoyle, J V Narlikar, Proc.
R. Soc. Lond. A, 282, 191 (1964)

\bibitem{Narlikar_Introduction to cosmology}J V Narlikar, An Introduction
to Cosmology, 3rd edition (Cambridge University Press, 2010), Chap.
9, p. 347

\bibitem{QSSC_arxiv review_Kragh}H Kragh, arXiv:1201.3449v1 {[}physics.hist-ph{]}
(2012)

\bibitem{QSSC deceleration_intergalactic dust}J V Narlikar, R G Vishwakarma
and G Burbidge, Pub. of the Astron. Soc. of the Pac., 114, 800, 1092
(2002)

\bibitem{Hoyle Burbidge Narlikar_Different approach to cosmology}F
Hoyle, G Burbidge and J V Narlikar, A different approach to cosmology
(Cambridge University Press, 2000), Chap. 16, p. 197\end{thebibliography}
\end{document}